\documentclass[aps, prl, superscriptaddress, nofootinbib, reprint, preprintnumbers]{revtex4-1}

\usepackage[utf8]{inputenc}
\usepackage[T1]{fontenc}
\usepackage{lmodern}
\usepackage{textcomp}
\usepackage{microtype}

\usepackage{mathtools}
\usepackage{physics}
\usepackage{units}
\usepackage{siunitx}

\newcommand{\scenario}[1]{\expandafter\newcommand\csname#1\endcsname[1]{\ensuremath{#1_{##1}}}}
\scenario{X}
\scenario{Y}

\DeclareMathOperator{\BR}{BR}

\usepackage{enumitem}
\setlist{noitemsep, topsep=2pt}

\usepackage[font={small,it}]{caption}
\usepackage{subcaption}

\usepackage{booktabs}

\usepackage{graphicx}
\graphicspath{{figs/}}
\newcommand{\image}[2][1]{\centering\includegraphics[width=#1\linewidth]{#2}}

\usepackage{tikz}
\usetikzlibrary{positioning}
\usetikzlibrary{shapes}
\DeclareUnicodeCharacter{FEFF}{~}

\usepackage[hidelinks]{hyperref}
\usepackage{xspace}
\newcommand{\ee}{\ensuremath{e^+e^-}\xspace}

\makeatletter
\g@addto@macro\bfseries{\boldmath}
\makeatother

\newcommand{\no}[1]{\textnumero~#1\xspace}

\newcommand{\fig}[1]{FIG.~\ref{fig:#1}}
\newcommand{\tab}[1]{TAB.~\ref{tab:#1}}

\usepackage{xcolor}

\begin{document}

\title{Novelty Detection Meets Collider Physics}

\author{Jan Hajer}
\affiliation{Institute for Advanced Studies, The Hong Kong University of Science and Technology, Clear Water Bay, Kowloon, Hong Kong S.A.R, P.R.China}
\affiliation{Centre for Cosmology, Particle Physics and Phenomenology, Université catholique de Louvain, Louvain-la-Neuve B-1348, Belgium}
\author{Ying-Ying Li}
\affiliation{Department of Physics, The Hong Kong University of Science and Technology, Clear Water Bay, Kowloon, Hong Kong S.A.R., P.R.China}
\affiliation{Kavli Institute for Theoretical Physics, University of California Santa Barbara, CA 93106--4030, USA}
\author{Tao Liu}
\author{He Wang}
\affiliation{Department of Physics, The Hong Kong University of Science and Technology, Clear Water Bay, Kowloon, Hong Kong S.A.R., P.R.China}

\preprint{CP3-18-51}

\begin{abstract}
Novelty detection is the machine learning task to recognize data, which belong to an unknown pattern. 
Complementary to supervised learning, it allows to analyze data model-independently.
We demonstrate the potential role of novelty detection in collider physics, using autoencoder-based deep neural network. 
Explicitly, we develop a set of density-based novelty evaluators, which are sensitive to the clustering of unknown-pattern testing data or new-physics signal events, for the design of detection algorithms. 
We also explore the influence of the known-pattern data fluctuations, arising from non-signal regions, on detection sensitivity.
Strategies to address it are proposed. 
The algorithms are applied to detecting fermionic di-top partner and resonant di-top productions at LHC,  and exotic Higgs decays of two specific modes at a future $e^+e^-$ collider. With parton-level analysis, we conclude that potentially the new-physics benchmarks can be recognized with high efficiency. 
\end{abstract}

\maketitle

\section{Introduction}

Since the early developments in the 1950's~\cite{Samuel59}, Machine Learning (ML) has evolved into a science addressing various \emph{big data} problems.
The techniques developed for ML, such as \emph{decision tree learning}~\cite{Quinlan86} and \emph{artificial neural networks} (ANN)~\cite{Peterson94}, allow to train computers in order to perform specific tasks usually deemed to be complex for handwoven algorithms.
For \emph{supervised learning}, the algorithm is first trained on labeled data, and then to classify testing data into the categories defined during training.
In contrast, in \emph{semi-supervised} and \emph{unsupervised learning}, where partially labeled or unlabeled data is provided, the algorithm is expected to find the relevant patterns unassistedly.

The last decade has seen a rapid progress in ML techniques, in particular the development of deep ANN.
A deep ANN is a multi-layered network of threshold units~\cite{LeCun15}.
Each unit computes only a simple nonlinear function of its inputs, which allows each layer to represent a certain level of relevant features.
Unlike traditional ML techniques (e.g.\ boosted decision trees) which rely heavily on expert-designed features in order to reduce the dimensionality of the problem, deep ANN automatically extract pertinent features from data, enabling data-mining without prior assumptions.
Fueled by vast amounts of big data and the fast development in training techniques and parallel computing architectures, modern deep learning systems have achieved major successes in computer vision~\cite{Krizhevsky12}, speech recognition~\cite{Hinton12}, natural language processing~\cite{Mikolov13}, and have recently emerged as a promising tool for scientific research~\cite{Zdeborova17, Carleo17, Carrasquilla17, Zhang18}, where the plethora of experimental data presents a challenge for insightful analysis.

High Energy Physics (HEP) is a big data science and has a long history of using supervised ML for data analysis.
Recently, pioneering works have demonstrated the capability of deep ANN in understanding jet substructure~\cite{Baldi16,Butter:2017cot, Larkoski:2017jix, Macaluso:2018tck} and the identification of particles~\cite{Baldi14} or even whole signal signatures (see e.g.~\cite{Cohen:2017exh}, where weakened supervised learning is applied).
However, the primary goal of the HEP experiments is to detect predicted or unpredicted physics Beyond the Standard Model (BSM) in order to establish the underlying fundamental laws of nature.
Despite its significant role in current data analysis, supervised ML techniques suffer from the model dependence introduced during training.
This problem can potentially be addressed by the semi-supervised/unsupervised techniques developed for \emph{novelty detection} (for a review see, e.g.~\cite{Pimentel14}).
Novelty detection is the ML task to recognize data belonging to an unknown pattern.
If being interpreted as novel signal, BSM physics could be detected without specifying an underlying theory during data analysis.
Hence, a combination of novelty detection and supervised ML may lay out a framework for the future HEP data analysis.

Some preliminary and at least partially related efforts have been made at jet~\cite{Metodiev:2017vrx, Andreassen:2018apy} and event~\cite{Aaltonen:2007dg, CMS:2008gya, ATLAS:2017irs, Kuusela:2011aa, Collins:2018epr, DAgnolo:2018cun} level. 
For novelty detection with given feature representation, its sensitivity depends crucially on the performance of novelty evaluators. Well-designed evaluators will allow to evaluate the data novelty efficiently and precisely.
As a matter of fact the design of novelty evaluators or the relevant test statistics defines the frontier of novelty detection~\cite{Pimentel14}.
In this letter, we propose a set of density-based novelty evaluators.
In contrast to traditional density-based ones, which only quantify isolation of testing data from the known patterns, the new novelty evaluators are sensitive to the clustering of testing data.
On this basis, we design algorithms for novelty detection using an autoencoder, which are subsequently applied for detecting several BSM benchmarks at LHC and future $e^+e^-$ colliders.

\section{Algorithms}

\begin{figure}
\includegraphics[width=0.6\columnwidth]{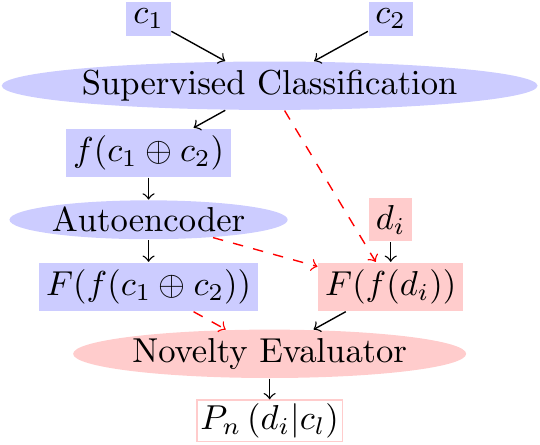}
\caption{%
Novelty detection algorithm.
The training and testing phases are marked in blue and red, respectively.
Datasets, algorithm and probabilities are indicated by rectangular, elliptic and plain nodes, respectively.
The information gathered during training and used for testing is marked by dashed red arrows.
For clarity we have limited the number of labeled known patterns $c_l$ to two. $d_i$ denotes testing data with known and unknown patterns.
}
\label{fig:algorithm}
\end{figure}

Novelty detection using a deep ANN can be separated into three steps:
1) feature learning, 2) dimensional reduction, 3) novelty evaluation.
During the first step the ANN is trained under supervision, using labeled known patterns.
The nodes of the trained ANN contain the information gathered for classification and constitute the feature space, which has typically a large dimension.
In order to reduce the sparse error and to improve the efficiency of the analysis, one removes the irrelevant features by \emph{dimensional reduction}, which can be implemented using an \emph{autoencoder}~\cite{Vincent08}.
An autoencoder is an ANN with identical number of nodes for input and output layers and fewer nodes for hidden layers.
Its loss-function measures the difference between input and output, defined as the reconstruction error $\norm{x - x^\prime}^2$.
Here $x$ and $x^\prime$ are the vectors of input and output nodes, respectively.
Hence the autoencoder learns unsupervised how to reconstruct its input.
This allows it to form a submanifold in the full feature space.
Afterwards, the novelty of testing data is evaluated, for the final significance analysis. 
The algorithm is shown in \fig{algorithm}.
For the HEP data analysis, the data with known and unknown patterns can be interpreted as SM background and BSM signal, respectively.

We generate Monte Carlo data using \texttt{Mad\-Graph5\_a\-MC@NLO}~\cite{Alwall:2014hca} and rely on \texttt{Keras}~\cite{Keras} (\texttt{TensorFlow}~\cite{tensorflow2015-whitepaper}-based) for the ANN construction.
For the \emph{supervised classification} of events with $n$ visible-particle four-momenta (which we internally normalise by $\unit[200]{GeV}$) and $l$ labeled patterns we use an ANN with $4 n$ input nodes, $l$ output nodes, and three hidden layers with 30, 30 and 10 nodes, respectively.
We use Nesterov's accelerated gradient descent optimizer~\cite{nesterov1983method} with a learning rate of 0.3, a learning momentum of 0.99 and a decay rate of $10^{-4}$.
The batch size is fixed to be 30 and the loss function is the categorical cross entropy~\cite{rubinstein1999cross, rubinstein2001combinatorial}.
The collection of all nodes constitute the feature space with dimension $m = 4 n + 30 + 30 + 10 + l$. This ensures that it contains the non-linear information learned from  classification.
We normalize the axes of the feature space to $[-1,1]$ and use $\tanh$ as activation function for the autoencoder.
Finally, an autoencoder consisting of five hidden layers with 40, 20, 8, 20 and 40 nodes, respectively, and a learning rate of 2.0 projects this feature space onto an eight-dimensional sub-space. 
We have checked that the results of all ANNs are stable against variations in the numbers of hidden layers and nodes.

\section{Novelty Evaluation}

\begin{figure}
\begin{subfigure}{0.49\linewidth}
\image{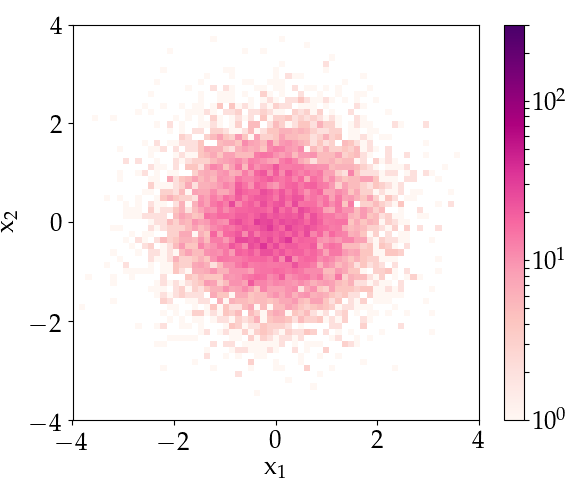}
\caption{Training data.}
\label{fig:toy_training}
\end{subfigure}
\hfill
\begin{subfigure}{0.49\linewidth}
\image{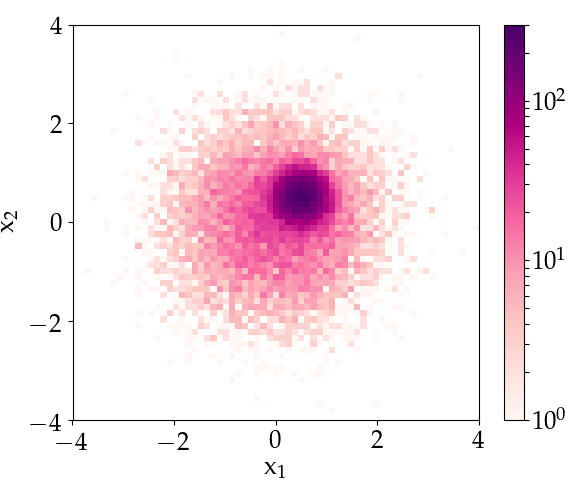}
\caption{Testing data.}
\label{fig:toy_testing}
\end{subfigure}
\begin{subfigure}{0.49\linewidth}
\image{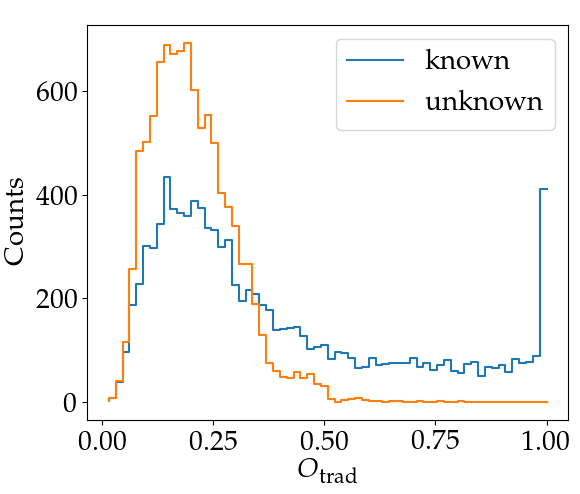}
\caption{${\mathcal O}_\text{trad}$ performance.}
\label{fig:ols1}
\end{subfigure}
\hfill
\begin{subfigure}{0.49\linewidth}
\image{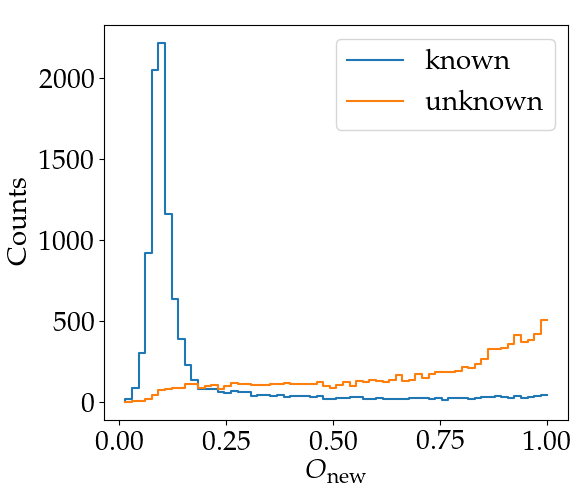}
\caption{${\mathcal O}_\text{new}$ performance.}
\label{fig:ols2}
\end{subfigure}
\caption{%
Comparison between traditional and new novelty evaluators.
The toy-data is shown in panels (\subref{fig:toy_training}) and (\subref{fig:toy_testing}), while the novelty response is given in (\subref{fig:ols1}) and (\subref{fig:ols2}).
}
\label{fig:nscore}
\end{figure}

\emph{Novelty evaluation} of testing data is a crucial step for novelty detection.
Various approaches have been developed in the past decades~\cite{Pimentel14}.
For non-time series data, one of the most popular approaches is density-based~\cite{Breunig00}, in which a Local Outlier Factor (LOF), i.e., the ratio of the local density of a given testing data and the local densities of its neighbors, is proposed as a novelty measure. Explicitly, this traditional measure is~\cite{Kriegel09,Socher13}
\begin{equation}
\Delta_\text{trad} = \frac{ d_\text{train} - \ev{d^\prime_\text{train}} }{ \ev{d^{\prime2}_\text{train}}^{\nicefrac{1}{2}}}
\ , \label{eq:traditional measure}
\end{equation}
here $d_\text{train}$ is the mean distance of a testing data to its $k$ nearest neighbors, $\ev{d^\prime_\text{train}}$ is the average of the mean distances defined for its $k$ nearest neighbors, and $\ev{d^{\prime2}_\text{train}}^{\nicefrac{1}{2}}$ is the standard deviation of the latter. 
The subscript of ``train'' indicates that all quantities are defined w.r.t the training dataset.
We calculate $\ev{d^{\prime2}_\text{train}}^{\nicefrac{1}{2}}$ using the method suggested in~\cite{Kriegel09,Socher13}.
The probabilistic novelty evaluator can be defined as the cumulative distribution function ${\mathcal O_{\rm trad}} = \frac12 \left( 1 + \erf{\frac{\Delta_{\rm trad}}{c\sqrt 2}}\right)$. Here $c$ is a normalization factor, defined as the root mean square of the measure values for all testing data.
This evaluator measures the isolation of testing data from training data.
A testing data located away from or at the tail of the training data distribution thus tends to be scored high by ${\mathcal O}_{\rm trad}$~\cite{Breunig00,Kriegel09}.

However, ${\mathcal O}_{\rm trad}$ is blind to the clustering of testing data which generically exists in the BSM datasets and may result in non-trivial structures such as resonance.  In order to utilize this feature, we introduce a measure:
\begin{equation}
\Delta_\text{new} = \frac{ d_\text{test}^{-m} - d_\text{train}^{-m} }{ d_\text{train}^{-\nicefrac{m}{2}}} \ ,
 \label{eq:new measure}
\end{equation}
with $m$ being the dimension of the feature space.
Here $d_\text{test}$ is the mean distance of the testing data to its $k$ nearest neighbors in the testing dataset, whereas $d_\text{train}$ is the SM prediction of the same, which can be approximately calculated using the training dataset.
This measure is reminiscent of the test statistic introduced in~\cite{Wang06,Dasu06}, where similar idea is employed for estimating the divergence of data distribution. 
As $\Delta_\text{new}$ is approximately $\propto \frac{S}{\sqrt{B}}$, with $S$ and $B$ being the numbers of signal and background events in a local bin with unit volume, this measure can be interpreted as the significance of discovery (up to a calibration constant) for this local bin.

\begin{figure}
\begin{subfigure}{0.49\linewidth}
\image{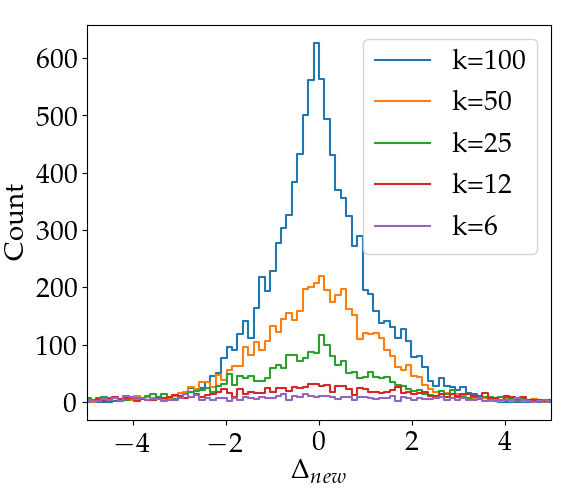}
\caption{$\Delta_\text{new}$ response.}
\label{fig:toy_K_Measure}
\end{subfigure}
\hfill
\begin{subfigure}{0.49\linewidth}
\image{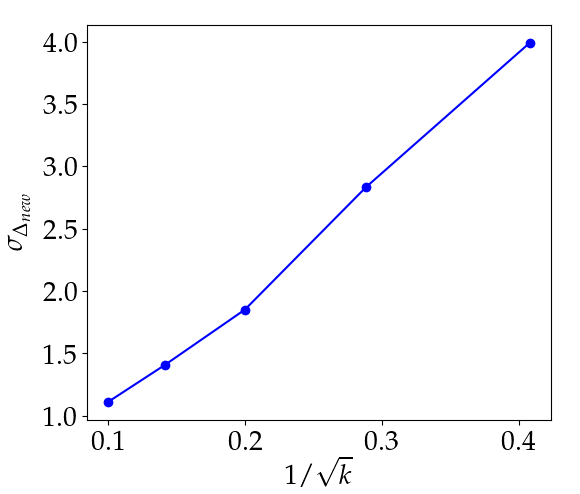}
\caption{CLT.}
\label{fig:toy_K_SD}
\end{subfigure}
\caption{%
Dependence of the $\Delta_\text{new}$ response on $k$, for the testing data with known patterns only. 
While the training dataset is composed of $50\,000$ points, the testing dataset consists of $10\,000$, $5\,000$, $2\,500$, $1\,250$ and 625 points, 
with $k$ scaling linearly as 100, 50, 25, 12 and 6, respectively. Both datasets are Gaussian. Panel~(\subref{fig:toy_K_Measure}) shows the $\Delta_\text{new}$ response in all cases.     Panel~(\subref{fig:toy_K_SD}) shows that its standard deviation $\sigma_{\Delta_\text{new}}$ scales linearly with $1/\sqrt k$ or $1/\sqrt L$, as predicted by the CLT.
}
\label{fig:fluc}
\end{figure}

 ${\mathcal O}_{\rm new}$ is defined in a similar way as ${\mathcal O}_{\rm trad}$ does. To compare the performance of ${\mathcal O}_{\rm trad}$ and ${\mathcal O}_{\rm new}$ in probing the clustering, we introduce a toy model, where the data resides in a two-dimensional space.
The known pattern is a Gaussian distribution centered around the origin $\mathcal N(\vec{0}, \mathbf I)$, while the unknown pattern is an overlapping narrow Gaussian distribution shifted away from the origin $\mathcal N((0.5, 0.5)^\mathsf T, 0.1 \mathbf I)$.
The training dataset consists of $10^4$ events with known pattern (cf.~\fig{toy_training}), while the testing dataset contains from each, known and unknown pattern, $10^4$ events (cf.~\fig{toy_testing}).
As shown in \fig{ols1} and \fig{ols2}, the clustering of the unknown-pattern data, although being hidden from ${\mathcal O}_{\rm trad}$, is picked-up by ${\mathcal O}_{\rm new}$.

The detection based on $\Delta_{\rm new}$ (or ${\mathcal O}_{\rm new}$) however may suffer from fluctuations of the known-pattern testing data in the non-signal regions, via the $1/d_{\rm test}^m$ term in Eq. (\ref{eq:new measure}). While $\Delta_\text{new}$ is expected to be zero if the data only consists of events with known patterns, the fluctuations result in non-zero values, since the measure picks up local data excess. This in essence is a kind of Look Elsewhere Effect (LEE). The fluctuations in $1/d_\text{train}^m$ on the other hand can be neglected, as long as the training dataset used for calculating $1/d_\text{train}^m$ is much larger than the testing one, with $k$ being properly scaled.

The influence of fluctuations on detection sensitivity can be compensated for as the luminosity $L$ increases, if $k$ scales with $L$. 
In this case more and more data are used to calculate $1/d_{\rm test}^m$ in the local bin which is barely changed. 
This compensation is approximately predicted by the Central Limit Theorem (CLT), which states in this context that the standard deviation of the $\Delta_\text{new}$ response scales with $1/\sqrt{k}$ or $1/\sqrt{L}$, for the testing data with known patterns only.
We show this in Fig~\ref{fig:fluc}, using the known-pattern Gaussian datasets defined before.  
Indeed, as the number of testing data increases, $\Delta_\text{new}$ becomes less and less sensitive to the fluctuations (see Fig.~\ref{fig:toy_K_Measure}). 

\begin{figure}
\begin{subfigure}{0.48\linewidth}
\image{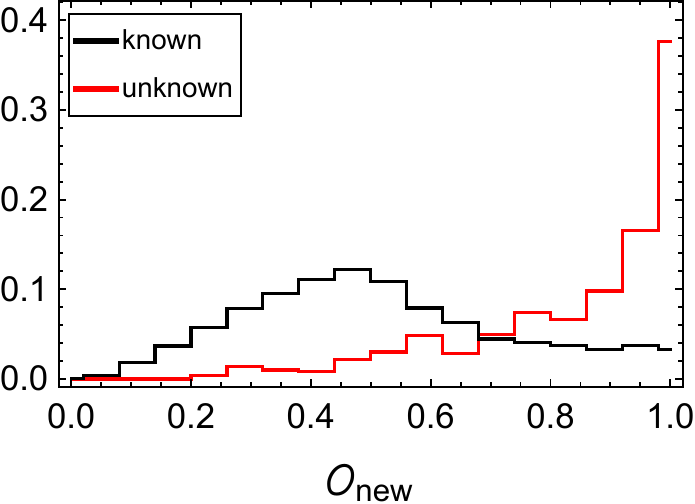}
\caption{New evaluator.}
\label{fig:toy_new}
\end{subfigure}
\hfill
\begin{subfigure}{0.48\linewidth}
\image{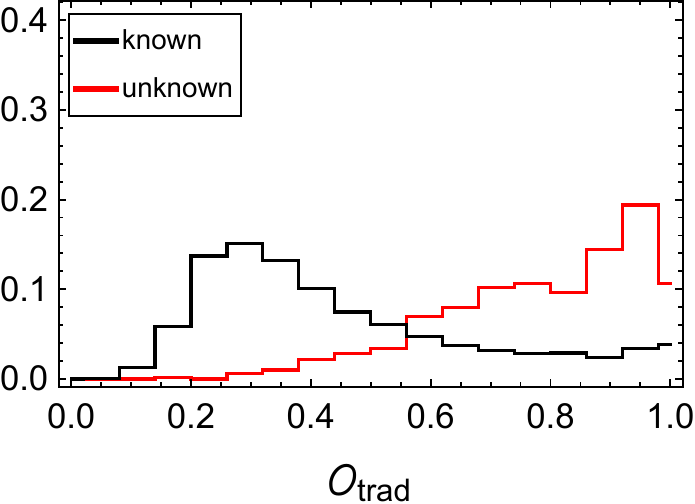}
\caption{Traditional evaluator.}
\label{fig:toy_trad}
\end{subfigure}
\begin{subfigure}{0.48\linewidth}
\image{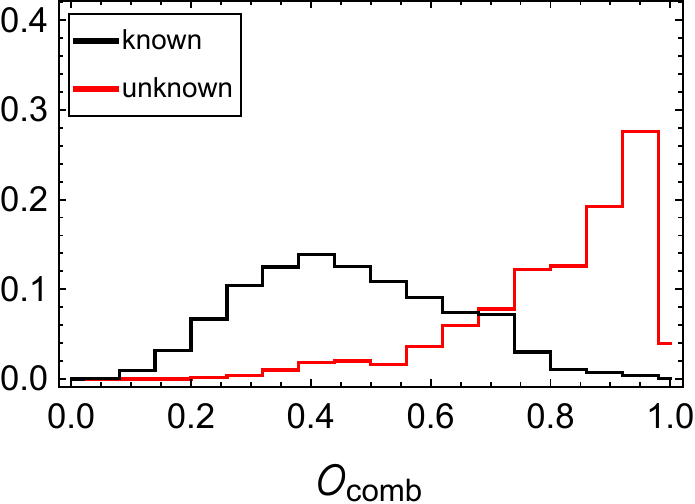}
\caption{Combined evaluator.}
\label{fig:toy_tot}
\end{subfigure}
\hfill
\begin{subfigure}{0.50\linewidth}
\image{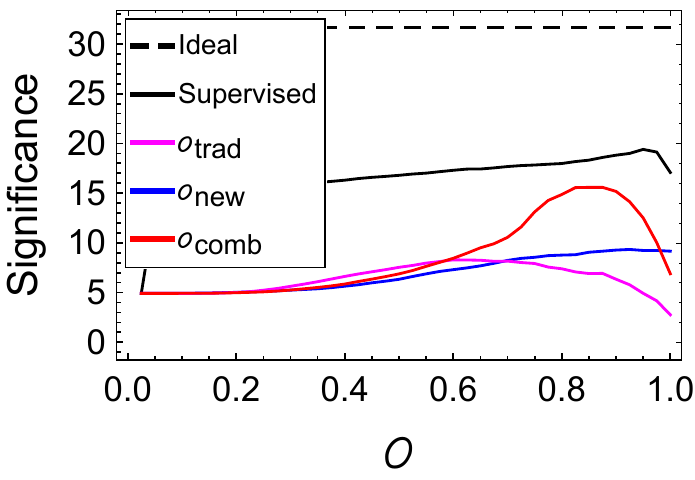}
\caption{Significance.}
\label{fig:toy_sensitivity}
\end{subfigure}
\caption{%
Normalized data responses to the novelty evaluators ${\mathcal O}_\text{trad}$~(\subref{fig:toy_trad}),  ${\mathcal O}_\text{new}$~(\subref{fig:toy_new}) and ${\mathcal O}_\text{comb}$~(\subref{fig:toy_tot}), and significance performance of these evaluators~(\subref{fig:toy_sensitivity}).
}
\label{fig:toy_sperformance}
\end{figure}

If the fluctuations are not fully compensated for by luminosity, the known-pattern testing data could still be scored high by $\Delta_\text{new}$, and hence diminish the detection sensitivity. This is often true if $S_{\rm tot} / B_{\rm tot}$ is small, as typically occurs in the analyses at LHC. To address this potential problem, we propose one more evaluator
\begin{equation}
{\mathcal O}_\text{comb} = \sqrt{ {\mathcal O}_\text{trad} {\mathcal O}_\text{new}} \ .
\end{equation}
This evaluator utilizes the fact that the known-pattern testing data with high ${\mathcal O}_\text{new}$ scores pretty often come from the high-density regions in the feature space, whereas such data are typically scored low by $\mathcal O_{\rm trad}$. As indicated in Fig.~\ref{fig:toy_sperformance}, ${\mathcal O}_\text{comb}$ performs very well in a typical case  where the known and unknown-pattern data distributions are partially overlapped, and many of the known-pattern data, especially the ones in the central region, are scored high by ${\mathcal O}_\text{new}$ due to the fluctuations. 
The known-pattern datasets used here are the same as before, containing $10^4$ events. The unknown pattern is defined as $\mathcal N((1.5, 1.5)^\mathsf T, 0.1 \mathbf I)$, with $\nicefrac{S_{\rm tot}}{B_{\rm tot}} = \nicefrac{1}{20}$. Indeed, many high-scoring data of known pattern in Fig.~\ref{fig:toy_new} are pushed to the low-scoring end in Fig.~\ref{fig:toy_tot}, due to the compensation of $\mathcal O_{\rm trad}$. 
This effect results in $\sim 50\%$ improvement in sensitivity, compared to the ones based on $\mathcal O_{\rm trad}$ or $\mathcal O_{\rm new}$ only.
Here (and similarly below) the significance is calculated against the known $+$ unknown-pattern hypothesis for testing data, using the Poisson-probability-based test statistic~\cite{Cowan:2010js}.

\section{Study on Benchmark Scenarios}

\begin{table}
\resizebox{\columnwidth}{!}{%
\begin{tabular}{|c|c|c|}
\hline
 & Parameter values & $\sigma (fb)$
\\ \hline \hline
    $X1$  & $m_T = m_{\overline T}$  \unit[1.2]{TeV},   $\BR(T \to W_l^+ b)= \unit[50]{\%}$  & 0.152
 \\ \hline $X2$  & $m_{Z^\prime} = \unit[3]{TeV}$,  $g_{Z^\prime} = g_Z$,    $\BR (Z^\prime \to \bar tt) = \unit[16.7]{\%}$ & 1.55
  \\   \hline  $Y1$  & $m_{N_1} = \frac{m_{N_2}}{9} = \frac{m_a}{4}= \unit[10]{GeV}$, $\BR(h \to \overline bb E_T^\text{miss})= \unit[1]{\%}$ & 0.108
 \\  \hline $Y2$  & $m_a = \unit[25]{GeV}$, $\BR(h \to \overline bb E_T^\text{miss})= \unit[1]{\%}$ & 0.053
\\ \hline
\end{tabular}
}
\caption{Parameter values and cross sections (after preselection) in the benchmark scenarios of BSM physics.}
\label{tab:benchmarks}
\end{table}

In order to illustrate their performance, we apply the algorithms designed above to two parton-level analyses, with two BSM benchmarks defined for each. Though being  unrealistic, it is sufficient for proof of concept. 

In the first analysis, we simulate the final state $\overline bbl^+l^-E_T^\text{miss}$ at the \unit[14]{TeV} LHC, with a luminosity of $\unit[3]{ab^{-1}}$.
We require exactly two bottom quarks with $p_T > \unit[20]{GeV}$ and two charged leptons ($e^\pm$ and $\mu^\pm$) with $p_T > \unit[10]{GeV}$.
The SM background stems mainly from
\begin{itemize}
\item \makebox[.4\linewidth][l]{$pp \to \bar t_l t_l$\ ,} $\sigma = \unit[11.5]{fb}$\ ,
\item \makebox[.4\linewidth][l]{$pp \to t_l \overline b W_l^\pm$\ ,} $\sigma = \unit[0.365]{fb}$\ ,
\item \makebox[.4\linewidth][l]{$pp \to Z_b Z_l$\ ,} $\sigma = \unit[0.0765]{fb}$\ .
\end{itemize}
Here the physical cross sections have been universally suppressed by a factor $2000$ for simplification.
The signal could arise from multiple BSM scenarios in this analysis. Here we consider:
\begin{description}
\item[\X1] $pp \to \overline T T \to W_l^+ W_l^- \overline bb$ where $\overline T$ and $T$ are fermionic top partners,
\item[\X2] $p p \to Z^\prime \to \bar t t$ where $Z^\prime$ is a new gauge boson.
\end{description}

In the second analysis, we simulate unpolarized $e^+e^- \to Zh$ production with the final state $\overline b b l^+ l^- E_T^\text{miss}$ at $\sqrt s =\unit[240]{GeV}$, with a luminosity of $\unit[5]{ab^{-1}}$.
We require exactly two bottom quarks with $p_T > \unit[10]{GeV}$ and  two charged leptons ($e^\pm$ and $\mu^\pm$) with $p_T > \unit[5]{GeV}$.
The SM background arises mainly from
\begin{itemize}
\item \makebox[.6\linewidth][l]{$e^+e^- \to hZ \to Z^*_\text{inv} Z_{\bar bb} l^+ l^-$\ ,} $\sigma = \unit[0.00686]{fb}$\ ,
\item \makebox[.6\linewidth][l]{$e^+e^- \to hZ \to Z_{\bar bb}^* Z_\text{inv} l^+ l^-$\ ,} $\sigma = \unit[0.00259]{fb}$\ .
\end{itemize}
For BSM scenarios, we consider two specific modes of exotic Higgs decay~\cite{Curtin:2013fra}:
\begin{description}
\item[\Y1] $h \to \widetilde \chi_1 \widetilde \chi_2 \to \widetilde \chi_1 \widetilde \chi_1 a$.
 This decay topology can arise from the nearly Peccei-Quinn symmetric limit in the NMSSM~\cite{Draper:2010ew,Huang:2013ima}, where $\widetilde \chi_2$ and $\widetilde \chi_1$ are bino- and singlino-like neutralinos, respectively, and $a$ is a light CP-odd scalar.
\item[\Y2] $h \to Z a $ in the 2HDM and the NMSSM~\cite{Curtin:2013fra}.
\end{description}
The parameter values and cross sections for the four benchmark scenarios are summarized in \tab{benchmarks}.

\begin{figure}
\begin{subfigure}{0.48\linewidth}
\image{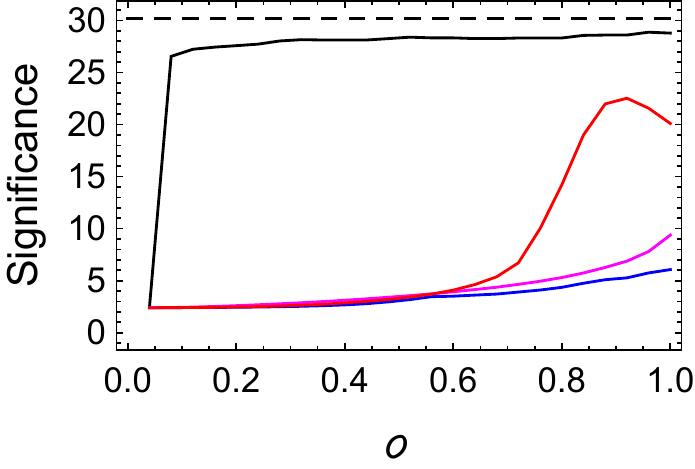}
\caption{Benchmark: $X_1$}
\label{fig:X1}
\end{subfigure}
\begin{subfigure}{0.48\linewidth}
\image{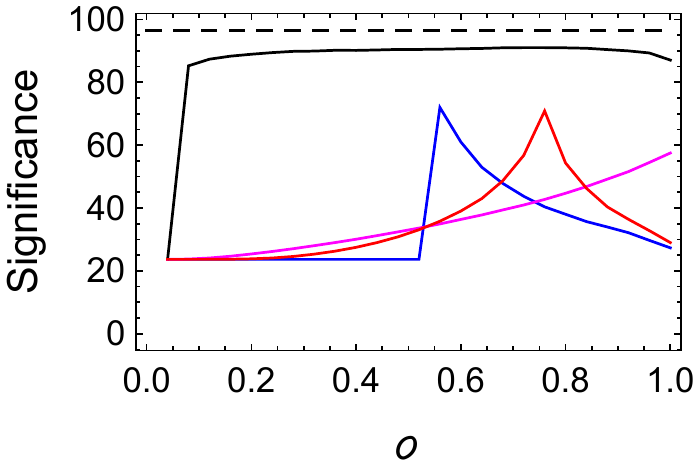}
\caption{Benchmark: $X_2$}
\label{fig:Y1}
\end{subfigure}
\begin{subfigure}{0.5\linewidth}
\image{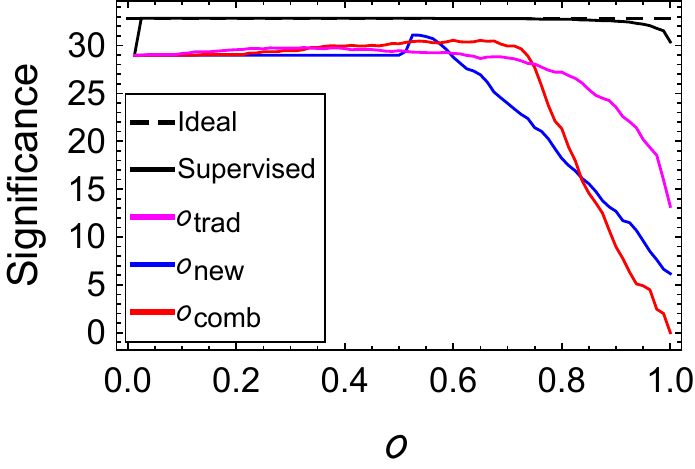}
\caption{Benchmark: $Y_1$}
\label{fig:X2}
\end{subfigure}
\begin{subfigure}{0.48\linewidth}
\image{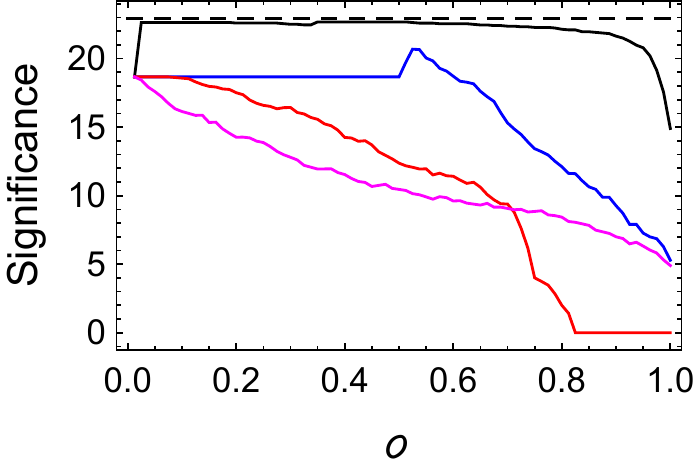}
\caption{Benchmark: $Y_2$}
\label{fig:Y2}
\end{subfigure}
\caption{
Significance performance of the novelty-detection algorithms.
}
\label{fig:sensitivity}
\end{figure}

The sensitivity performance of the algorithms is presented in \fig{sensitivity}.
In each panel, we show one curve in the ``Ideal'' case (assuming 100\% signal efficiency and background rejection) and one curve with supervised learning as the references for  performance evaluation.
In the first analysis, the toy model discussed above precisely mimics what happens in benchmark \X1.
In this case, the BSM signal and the SM data are partially overlapped in the feature space. 
Many of the SM data in the non-signal regions have a strong ${\mathcal O}_\text{new}$ response, due to fluctuations, and hence diminish the detection sensitivity. 
However, with the ${\mathcal O}_\text{trad}$ compensation, sizable improvement in sensitivity is achieved. 
As shown in Fig.~\ref{fig:X1}, the sensitivity is approximately doubled using ${\mathcal O}_\text{comb}$, compared to the ones using ${\mathcal O}_\text{new}$ or ${\mathcal O}_\text{trad}$ only.
For benchmark \X2, $S_{\rm tot} / B_{\rm tot}$ is about one order larger than that in benchmark \X1, as indicated in \tab{benchmarks}. This tends to enhance the ${\mathcal O}_\text{new}$ response of the signal, compared to the SM data, and hence results in comparable sensitivities for the analyses based on ${\mathcal O}_\text{trad}$, ${\mathcal O}_\text{new}$ and ${\mathcal O}_\text{comb}$, respectively.
For the benchmarks \Y1 and \Y2 in the second analysis, the fluctuation effect on ${\mathcal O}_\text{new}$ is negligibly small, due to $S_{\rm tot} / B_{\rm tot} > 1$ (typical for the analyses at \ee collider), while the known- and unknown-pattern data distributions are not fully separated, hence limiting the efficiency of ${\mathcal O}_\text{trad}$. This results in a  sensitivity performance for ${\mathcal O}_\text{new}$ which is universally better than the others.

\section{Summary and Discussion}

In this letter, we proposed a set of density-based novelty evaluators, ${\mathcal O}_\text{new}$ and ${\mathcal O}_\text{comb}$, which are sensitive to the clustering of the unknown-pattern testing data, for novelty detection in the HEP data analysis. 
These evaluators allow to design the algorithms with broad applications in detecting BSM physics. 
They can be also applied to measuring the SM processes yet to be discovered, if we interpret them as ``novel'' events. 
As these algorithms are designed using only general assumptions their application could be extended to other big-data domains as well.

This study could be generalized in multiple directions.
We have focused on developing the algorithms for novelty detection in HEP, using parton-level analysis to demonstrate their sensitivity performance. 
To fill up the gap between the concept and its application to real data analysis, hadron-level analysis is definitely needed. 
In addition, the algorithms could be improved in several aspects. First, the feature selection in the ANN training process might be not yet fully optimized. The features learned from classification of data with labeled known patterns are likely to be sub-optimal for enhancing the isolation or clustering of the unknown-pattern data. Nevertheless, we may introduce dynamical ML or some feedback mechanisms using the testing dataset, to reinforce the learning of the unknown-pattern features. Second, the distance definition of data depends on the geometry of the feature space. We adopted the Euclidean geometry for simplicity, but it is worthwhile to explore the other possibilities. Third, the amount of memory and time needed to implement $O_{trad}$ increases rapidly with the data size and dimension, which renders $O_{trad}$ not very efficient for large dataset. Ways of accelerating the calculation might be needed. 
More than that, we would extend the performance analysis of the algorithms to other BSM scenarios, e.g., the ones with interference between the known and unknown patterns, or non-trivial data clusters such as a dip~\cite{Dicus:1994bm}.
Although it is beyond the scope of this study, at last we mention that, a full analysis of the systematic and theoretical uncertainties is absent (for recent effort partially addressing this see~\cite{Englert:2018mbw}). 
We leave these topics to a future study.

\paragraph{Note added}

While this letter was being finalized, \cite{DeSimone:2018efk} appeared. 
Both the novelty evaluators proposed here and the test statistic defined in~\cite{DeSimone:2018efk} (as well as the one developed in~\cite{DAgnolo:2018cun} recently)
are able to measure the clustering of testing data with unknown pattern.
We would like to stress that we developed this project and the relevant ideas independently.
Particularly, two significant differences exist between them. 
First, unlike the test statistic in~\cite{DAgnolo:2018cun,DeSimone:2018efk} which measures the divergence of the testing dataset from the training dataset, the evaluators proposed quantify the novelty of individual testing data. Such a design difference enables the evaluators to probe the fine/differential structure of the clustering such as peak-dip (a famous BSM example can be found in~\cite{Dicus:1994bm}) more efficiently.   
Second, as the LEE could be a severe problem for novelty detection at Hadron colliders, we explored how to diminish its influences on detection sensitivity (in relation to this, ${\mathcal O}_{\rm comb}$ was designed). This was not developed in~\cite{DAgnolo:2018cun,DeSimone:2018efk}.

\subsection{Acknowledgments}

\begin{acknowledgments}
We would greatly thank Prof.\ Michael Wong, our colleague at the HKUST, for highly valuable discussions on the novelty evaluators and the ANN algorithms which were proposed in this letter.
T.~Liu would thank Huai-Ke Guo for discussions on CLT in this context during the MITP (the Mainz Institute for Theoretical Physics) workshop ``Probing Baryogenesis via LHC and Gravitational Wave Signatures'', June, 2018.
We would thank the experimental colleagues Aurelio Juste, Kirill Prokofiev and Junjie Zhu for reading the manuscript and raising valuable comments.
We would also thank Lian-Tao Wang and Zhen Liu for general discussions on this idea at an early stage.
J.~Hajer is partly supported by the General Research Fund (GRF) under Grant \no{16304315}.
Y.~Y.~Li would thank the Kavli Institute for Theoretical Physics, where most of her work was done, for the award of the graduate fellowship which was provided by Simons Foundation under Grant \no{216179} and Gordon and Betty Moore Foundation under Grant \no{4310}.
This research was also supported in part by the National Science Foundation under Grant \no{PHY-1748958}.
T.~Liu is jointly supported by the GRF under Grant \no{16312716 and 16302117}.
The GRF is issued by the Research Grants Council of Hong Kong S.A.R.
He would also thank the MITP for its hospitality, where part of his work was done.
\end{acknowledgments}

\bibliography{neuralnetwork}

\end{document}